# A MULTIPLICATIVE MASKING METHOD FOR PRESERVING THE SKEWNESS OF THE ORIGINAL MICRO-RECORDS

Nicolas Ruiz[1]

OECD

## ABSTRACT

Masking methods for the safe dissemination of microdata consist of distorting the original data while preserving a pre-defined set of statistical properties in the microdata. For continuous variables, available methodologies rely essentially on matrix masking and in particular on adding noise to the original values, using more or less refined procedures depending on the extent of information that one seeks to preserve. Almost all of these methods make use of the critical assumption that the original datasets follow a normal distribution and/or that the noise has such a distribution. This assumption is, however, restrictive in the sense that few variables follow empirically a Gaussian pattern: the distribution of household income, for example, is positively skewed, and this skewness is essential information that has to be considered and preserved. This paper addresses these issues by presenting a simple multiplicative masking method that preserves skewness of the original data while offering a sufficient level of disclosure risk control. Numerical examples are provided, leading to the suggestion that this method could be well-suited for the dissemination of a broad range of microdata, including those based on administrative and business records.

---

[1] Contact : nicolas.ruiz@oecd.org

**Introduction**

Microdata are individual records coming from surveys or administrative registers. Due to their nature, they provide a rich amount of information that can inform statistical and policy analysis.. However, this wealth of information is often untapped due to the legal obligations that National Statistical Offices (NSOs) and other governmental institutions face to protect the confidentiality of their respondents. Such requirements shape the dissemination policy of microdata at national and international levels. The issue is how to ensure a sufficient level of data protection to meet data producers' concerns in terms of legal and ethical requirements while offering to users a reasonable richness of information. To solve this tension, several solutions are available. These include providing access to microdata through a controlled environment such as data centres, safe remote access, providing interval data or tabulations rather than data points, or modifying the individual records before public release by using statistical disclosure control techniques.

Over the last decade, the role of micro-data has changed from being the preserve of NSOs and government departments to being a vital tool for analysts trying to understand both social and economic phenomena. These new uses of micro-data confront providers of official statistics with a new range of questions. The OECD has witnessed this change in microdata use and has become more actively involved in the subject. The overriding principle followed by national statistical offices when providing access to this information is that, in all instances, the confidentiality of individual responses should be preserved. The need to reassure microdata providers that their confidentiality constraints would be met has shaped the way the OECD has approached the issues of access to these data. Three themes are at the core of such access:

1. The development of an internationally-harmonised nomenclature and coding for microdata variables.

2. The development of an international Statistical Disclosure Control technique.

3. The development of an IT infrastructure to allow remote access in a secure environment.

This paper addresses the second theme. It proposes a harmonised statistical disclosure technique that is based on experiences from individual countries. This technique has the potential to lead to internationally comparable results as well as to ensure providers of official statistics that they retain control over the disclosure risk of their micro-data. While the initial investigation of disclosure techniques undertaken by the OECD focused on Labour Force Survey microdata (OECD 2010) the tool presented here has a true potential to be applied to a range of other domains where micro-data are potentially highly concentrated in the tails of the distribution, as in the case of business or household income and wealth data.

**Statistical Disclosure Control: what is it?**

Statistical Disclosure Control (SDC) consists in the set of numerical tools that enhances the level of confidentiality of any given micro-record while preserving to a lesser or greater extent its level of information (Hundepool and al., 2010, for an authoritative survey). While standards are still missing for the use of SDC in an integrated and coherent framework both at the national and international levels, some techniques are worth looking at due to their tractability and their performance regarding the trade-off between confidentiality and information. Among them, data perturbation has gained considerable attention in the literature.

Data perturbation involves distortions of the original datasets such that unique combinations of original values disappear and new ones are created. This perturbation is made to preserve statistical confidentiality. At the same time, statistical properties (more specifically a selected sub-set of them) of the original data are preserved, or do not differ significantly. The selection of suitable perturbation methods requires choosing those that will maximise statistical information while minimizing disclosure risk. No dominant method exists, in the sense that the type of statistical information preserved differs among the different techniques available and for different associated levels of disclosure risk.

The general approach of data perturbation consists in the matching of the original data with random noise terms in a non-reversible way, i.e. the data user cannot recover the original values from the perturbed ones. This can be performed in various ways, from a simple additive structure to non-linear transformations, applicable to both categorical and numerical variables. However, most of the perturbation techniques focus on continuous variables and so will the methodology presented in this paper.

In practice, popular perturbation techniques (Brand, 2002; Burridge, 2003; Muralidhar and al., 2005) use an additive structure for noise application, where error terms are randomly draw from a normal distribution, the latter being data-dependently parameterized in such a way that the resulting distribution of the perturbed values have the same first and second order moments as those in the original data. As information on these two moments is sufficient to fully identify a normal distribution, this implies that if the original values follow themselves a normal law then the original and the perturbed values will have exactly the same distribution. The loss of statistical information is thus low, in the sense that only the values of the data points of the underlying distribution are altered but not their overall shape. Such a high degree of preservation is made possible by the use of the Gaussian framework. Apart from its peculiar properties, the choice of additive noise methods is motivated by the fact that normality underlies many statistical and econometric tools, extending thus the usefulness and audience for these techniques.

Additive noise methods have, nonetheless, some drawbacks. The most obvious and crucial is the amount of information that is lost when the original data do not follow a normal law. In this case, analysis performed on perturbed data could produce quite different results from those performed on the original set. In particular, the Gaussian framework implies a strong assumption of symmetry in the original distribution. Clearly for numerous economic variables, this assumption is too strong to be tenable.

In fact, microdata often exhibit positively skewed distributions, as in the case of household income and wealth. Recent studies relying on a growing stream of research on income inequality (such as Piketty and al., 2003; and Atkinson and al., 2010), have pointed out the fact that in most developed countries top incomes contribute disproportionately to the overall level of income inequality in an country. As a result, skewness matters, and perturbation methodologies preserving it are of central interest for statistical disclosure control, despite its lack of treatment in the literature (see Muralhidar and al., 1995 for an attempt). In such cases, lognormal distributions appear to display a reasonable approximation for a large range of economic variables (Kleber and al., 2003; Lydall, 1966). As such, Gaussian perturbation methods would be of limited utility when applied to such distributions for at least two reasons:

- First, the sum of skewed and non-skewed distribution provides an identifiable distribution in very rare cases (Gao and al., 2009; Krishnamoorthy, 2006). Thus perturbed datasets will, in most cases, follow unknown and unidentifiable distributions.

- Second, as the presence of observations far from the mean leads to a skewed distribution, it follows that adding noise drawn from a normal distribution to those observations will only weakly perturb them. As an example, very large firms in business surveys will be re-submitted to high disclosure risk after perturbation, hence raising the issue of protection and confidentiality.

This paper presents a new multiplicative masking method that preserves positive skewness of the original data based on the lognormal distributions. This method allows users to generate perturbed data that are similar to the original data to a degree that is selected by the user. The methodology preserves confidentiality constraints in particular for observations away from the mean, by swapping them in the sample during the perturbation process. The methodology will be presented in the next section, after having first described the features of a general additive Gaussian method based on Muralhidar et al., 2008. The fourth section proposes numerical validation. The last section concludes.

**Methodology**

Described in this section is the proposed methodology for preservation of asymmetric distribution based on the identification of sufficiency conditions for lognormal distribution. To fully appraise the departure from additive Gaussian methods, we first describe the latter using the recent methodology of Muralhidar & al. (2008), showing how it is possible to generate perturbed data that preserves the distribution of the original dataset but where data points have a selectable degree of similarity.

*The Muralhidar-Sarathy hybrid generator*

Let's assume that X is a confidential variable that we want to perturb, and that S is a non-confidential[2] or a key variable with a low level of identification risk. Without loss of generality, it is assumed that the means of X and S are equal to zero. Let $\sigma_{XX}^2$, $\sigma_{SS}^2$ and $\sigma_{SX}^2$ be respectively the variance of X, S and the covariance between X and S. We will denote by Y the perturbed value of X generated by the following equation (where $y_i, x_i, s_i$ $\forall i=1,...,n$ are the values of Y, X and S variables for the ith respondent in the dataset):

$$y_i = \left[ (1-\alpha)\frac{1}{n}\sum_{i=1}^{n} x_i - \beta \frac{1}{n}\sum_{i=1}^{n} s_i \right] + \alpha x_i + \beta s_i + u_i \quad \forall i=1,...,n$$

α and β are coefficients and $u_i$ is a random term generated from a normal distribution $N(0, \sigma_{uu}^2)$, satisfying $\frac{1}{n}\sum_{i=1}^{n} x_i u_i = \frac{1}{n}\sum_{i=1}^{n} s_i u_i = 0$ ($x_i$ and $s_i$ are orthogonal to $u_i$). This equation shows that α can be interpreted as a similarity parameter between Y and X. When α=0, X and Y are completely dissimilar. For α=1 Y equals X and no perturbation is added. Thus, the choice of α allows the user (e.g. national statistical offices in the case of records from official sources) to control for the degree of similarity between the original and the perturbed variable that will be disseminated.

The conversion of X into Y through the preceding equation adds 'noise' to the original variable X. In fact, it is easy to verify that $E(y_i) = \frac{1}{n}\sum_{i=1}^{n} x_i$ and thus that X and Y will have the same expectation:

---

[2] Confidential variables are variables containing sensitive information that has to be protected from disclosure risk; non-confidential variables are variables for which disclosure does not raise any confidentiality issue.

the first moment of X's distribution is then preserved. To preserve the second moment, the following condition must be satisfied:

$$\sigma^2_{XX} = \sigma^2_{YY} = E\left[(\alpha x_i + \beta s_i + u_i)(\alpha x_i + \beta s_i + u_i)\right]$$
$$= \alpha^2 \sigma^2_{XX} + \beta^2 \sigma^2_{SS} + \sigma^2_{uu} + 2\alpha\beta\sigma^2_{SX}$$

Finally in order to preserve the covariance between the confidential and non-confidential variables, the following equation must also hold:

$$\sigma^2_{SX} = \sigma^2_{SY} = \alpha\sigma^2_{SX} + \beta\sigma^2_{SS} \text{ (as } \frac{1}{n}\sum_{i=1}^{n} s_i u_i = 0\text{)}$$
$$\Leftrightarrow \beta = (1-\alpha)\frac{\sigma^2_{SX}}{\sigma^2_{SS}}$$

Combining the two preceding equations above, we obtain the following restriction for $\sigma^2_{uu}$:

$$\sigma^2_{uu} = (1-\alpha^2)\left[\sigma^2_{XX} - \frac{(\sigma^2_{SX})^2}{\sigma^2_{SS}}\right]$$

The term $\left[\sigma^2_{XX} - \frac{(\sigma^2_{SX})^2}{\sigma^2_{SS}}\right]$ is always greater than or equal to zero. Thus, the necessary and sufficient condition to have $\sigma^2_{uu} > 0$ is that $-1 \leq \alpha \leq 1$. As a negative α induces a negative correlation between the original and the perturbed value, this case is ignored in the following, i.e. we will focus only on $0 \leq \alpha \leq 1$ to fulfil the above restrictions.

When α is set to 1, X=Y and no perturbation is added; when α=0, Y is not a function of the (confidential) value X but only of the non-confidential variable S and of an error term. The intermediary cases where 0< α<1 create therefore a hybrid dataset, as the released variable is a combination of its original value, of the non-confidential variable S and of a noise term. Through this method, users can thus choose to which extent they want to protect their initial release. This procedure is perfectly secure in the sense that no reverse engineering is possible as the hybridation is performed using a random draw for $u_i$. A direct consequence of this algorithm is that users can choose to communicate transparently their chosen degree of 'dissimilarity': in other terms, knowledge of α provides access to the value of $\sigma^2_{uu}$ but not to the $u_i$ values themselves (although maintaining the confidentiality of α does provide an additional security gate).

While it can be argued that this method implies significant information loss, in fact statistical information is preserved to a greater extent than with other approaches (such as those described by Fuller, 1993). In particular, the Muralidhar-Sarathy method preserves the first two moments of variable X's distribution, these moments being the necessary and sufficient conditions for the identification of a normal distribution; it follows that if the distribution of X is normal, then Y will have exactly the same distribution as the original, undisclosed variable. Moreover, by using a non-confidential variable in the perturbation process, this method allows preserving the covariance between the confidential variable X and the non-confidential variable S.

As appealing as this framework is, it relies nonetheless on the pivotal normality assumption. Normality underlies many statistical analyses commonly used (such as regressions and hypothesis tests), and assures that analysis based on the masked data will lead to the same results that one would have obtained with the original data – but with the advantage that the secure environment avoids disclosure risks. But the methodology is rather limitative if, rather than being interested in using the data for econometrics and inference, users are interested in the intrinsic features of the distribution, e.g. to compute descriptive statistics such as fractiles or measures of dispersions. In this case, perturbation using additive Gaussian noise loses its usefulness as additional features of the original distribution have to be preserved, in particular skewness which conveys substantial and relevant information.

*A sufficient multiplicative masking method for lognormal distributions*

Using the same notations as before, we let X follow a lognormal distribution with parameters $\mu_X > 0$ and $\sigma_{XX}^2$ :

$$X \mapsto LN(\mu_X; \sigma_{XX}^2)$$

Where, by definition of a lognormal distribution, $\mu_X = \frac{1}{n}\sum_{i=1}^{n} \ln x_i$ and $\sigma_{XX}^2 = \frac{1}{n}\sum_{i=1}^{n}(\ln x_i - \mu_X)^2$. The first and second order moments of X are thus respectively:

$$E(X) = \exp\left(\mu_X + \frac{\sigma_{XX}^2}{2}\right)$$

$$\text{and } V(X) = \left[\exp(\sigma_{XX}^2) - 1\right]\exp(2\mu_X + \sigma_{XX}^2)$$

The same assumptions apply for the perturbation U, assumed to be independent of X and with parameters $\mu_U = \frac{1}{n}\sum_{i=1}^{n} \ln u_i > 0$ and $\sigma_{UU}^2 = \frac{1}{n}\sum_{i=1}^{n}(\ln u_i - \mu_u)^2$ :

$$U \mapsto LN(\mu_U; \sigma_{UU}^2)$$

with $E(U) = \exp\left(\mu_U + \frac{\sigma_{UU}^2}{2}\right)$ and $V(U) = \left[\exp(\sigma_{UU}^2) - 1\right]\exp(2\mu_U + \sigma_{UU}^2)$

The perturbed value of X, Y is generated through the following equation, a homothetic Cobb-Douglas function:

$$Y = X^\alpha U^{1-\alpha} \text{ with } 0 \leq \alpha \leq 1$$

As for the Muralhidar-Sarathy hybrid generator, α is also a similarity parameter: when α is set to 1, X=Y and no perturbation is generated; when α=0, Y is not a function of the confidential value X but only of the lognormal noise. The intermediary cases 0< α<1 create convex combinations of confidential values and noise.

The properties of lognormal distribution (Krishnamoorthy, 2006) ensure that the α power distribution of X also follows a lognormal law:

$$X^{\alpha} \mapsto LN(\alpha\mu_X; \alpha^2 \sigma_{XX}^2)$$

and the same applies for the 1- α power of U:

$$U^{1-\alpha} \mapsto LN((1-\alpha)\mu_U; (1-\alpha)^2 \sigma_{UU}^2)$$

Given independency of U and X, Y has thus the following distribution:

$$Y \mapsto LN(\alpha\mu_X + (1-\alpha)\mu_U; \alpha^2 \sigma_{XX}^2 + (1-\alpha)^2 \sigma_{UU}^2)$$

with the associated two first moments being: $E(Y) = \exp\left(\alpha\mu_X + (1-\alpha)\mu_U + \dfrac{\alpha^2 \sigma_{XX}^2 + (1-\alpha)^2 \sigma_{UU}^2}{2}\right)$ and $V(Y) = \left[\exp\left(\alpha^2 \sigma_{XX}^2 + (1-\alpha)^2 \sigma_{UU}^2\right) - 1\right] \exp\left[2\left(\alpha\mu_X + (1-\alpha)\mu_U\right) + \alpha^2 \sigma_{XX}^2 + (1-\alpha)^2 \sigma_{UU}^2\right]$.

We can now derive the necessary and sufficient conditions that will ensure that Y has the same distribution as X. Unlike the additive framework, we cannot proceed by preserving the first two moments of Y. More generally any set of k-order moments with k≥1 is not isomorphic to any lognormal law: we can in fact always find other laws (lognormal or not) that have the same moments. To achieve sufficiency we have to consider the logarithmic transformation of Y:

$$\ln Y \mapsto N(\alpha\mu_X + (1-\alpha)\mu_U; \alpha^2 \sigma_{XX}^2 + (1-\alpha)^2 \sigma_{UU}^2)$$

Being now in a Gaussian case, we can derive conditions for the first two moments:

$$\alpha\mu_X + (1-\alpha)\mu_U = \mu_X \Leftrightarrow \mu_X = \mu_U$$

$$\alpha^2 \sigma_{XX}^2 + (1-\alpha)^2 \sigma_{UU}^2 = \sigma_{XX}^2 \Leftrightarrow \sigma_{UU}^2 = \dfrac{1-\alpha^2}{(1-\alpha)^2} \sigma_{XX}^2$$

As $\sigma_{UU}^2 \geq 0$, we also have $1-\alpha^2 \geq 0$ and thus $0 \leq \alpha \leq 1$, confirming $\alpha$ as a well-defined similarity parameter. Using the sufficiency conditions at the logarithmic level and exponentiating $\ln Y$, we find that U must have the following lognormal distribution:

$$U \mapsto LN(\mu_X; \dfrac{1-\alpha^2}{(1-\alpha)^2} \sigma_{XX}^2)$$

As exponentiation establish a one to one correspondence (i.e. it is a bijective mapping), the sufficiency conditions at the logarithmic scale ensure sufficiency at the original variable scale. Thus this perturbation method preserves the features of the original distribution including its skewness, but allows the similarity of data points to be selected. As shown in the following section, this method is also confidentiality efficient in the sense that disclosure risk remains low, in particular for observations far from the mean.

**Numerical validation**

Methods for statistical disclosure control cannot be fully appraised without experimental validation[3]. We simulated a vector consisting of one thousand data points drawn from a lognormal distribution with parameters 4 and 2, i.e. a deliberately highly skewed distribution. Figure 1 shows the density of the original distribution.

**Figure 1. Density of original data**

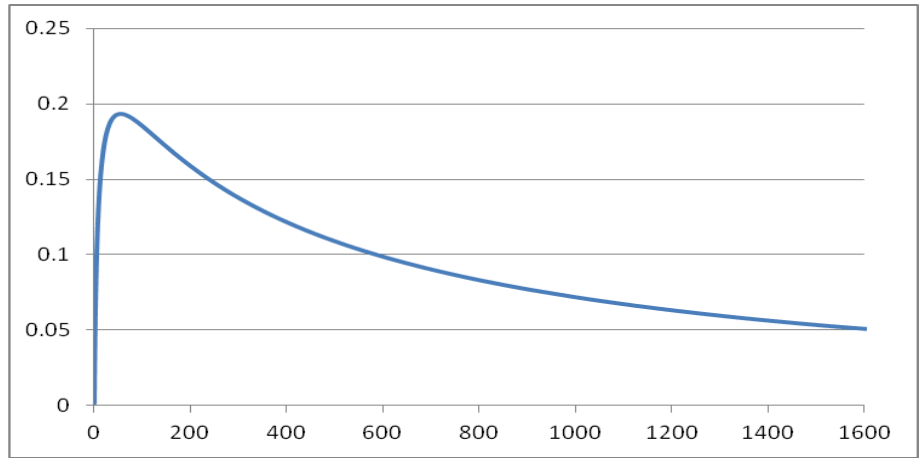

When α=0.9, the distribution of the perturbed data matches exactly that of the original data: as one can see in Figure 2, the density of the former is strictly identical to the latter.

**Figure 2. Density of perturbed data with α =0.9**

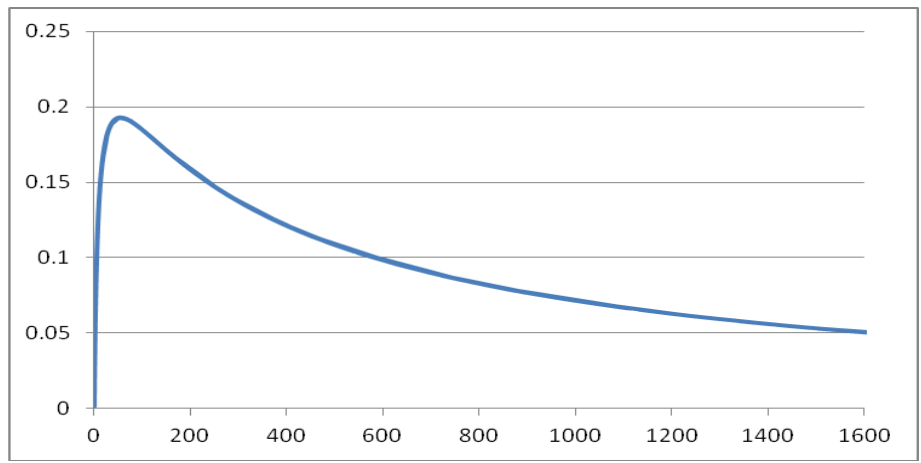

As derived in the previous section, perturbed distributions will remain the same as the original one for $0 \leq \alpha \leq 1$. Thus, the multiplicative masking method preserves the initial data structure. Nonetheless, data points are altered in an interesting way, in particular for confidentiality purpose. Figure 3 depicts the changes that occur in the absolute values for each point (ranked in ascending order on the x-axis according to their original values).

---

[3] Program and the dataset are available upon request to the author.

**Figure 3.  Absolute differences between original and perturbed data for α =0.999**

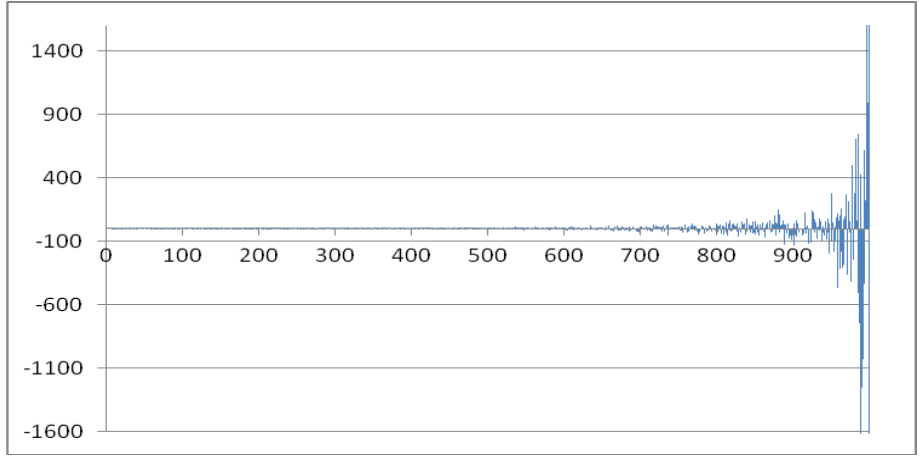

One immediately sees that, for a small value of the dissimilarity parameter, most of the data points that are close to the mean are very close to the original values while, due to the multiplicative structure used, values that are far away from the population mean are substantially altered. And as high values are those where disclosure risk is higher, this pattern of perturbation is the one most appropriate. For lower values of $\alpha$, and thus greater dissimilarity, perturbations start to spread along the distribution, from the upper to the lower tail as can be seen in Figures 4, 5 and 6.

**Figure 4.  Absolute differences between original and perturbed data for α =0.95**

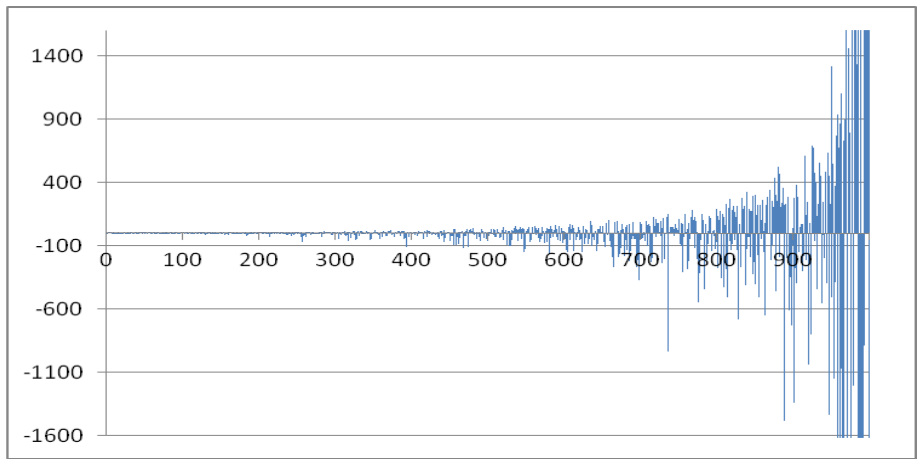

**Figure 5. Absolute differences between original and perturbed data for α =0.9**

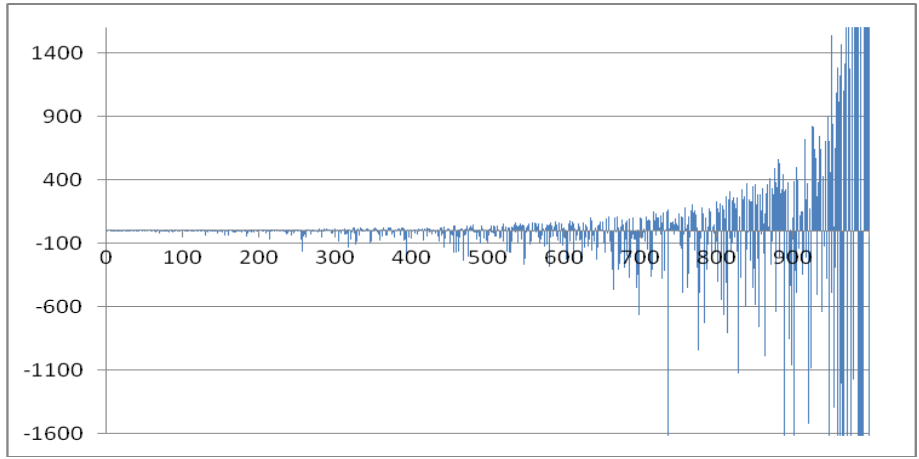

**Figure 6. Absolute differences between original and perturbed data for α =0.7**

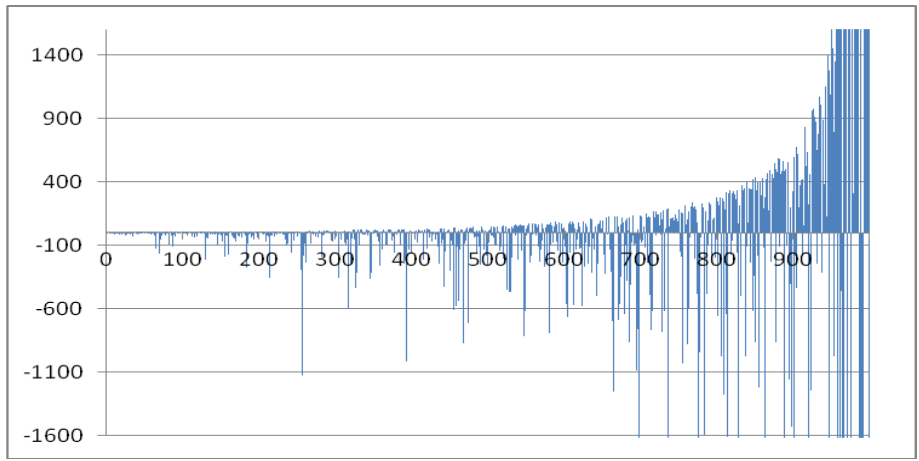

As perturbations can both reduce and increase values of different data points, the ranking of data points is likely to change during the process, thus increasing data protection against disclosure risk (in particular observations away from the mean could now become closed to it and conversely). As shown in Figures 7 and 8, the more dissimilarity is introduced, the more swaps occur in the data ranking.

**Figure 7. Initial vs. perturbed ranks for α =0.95**

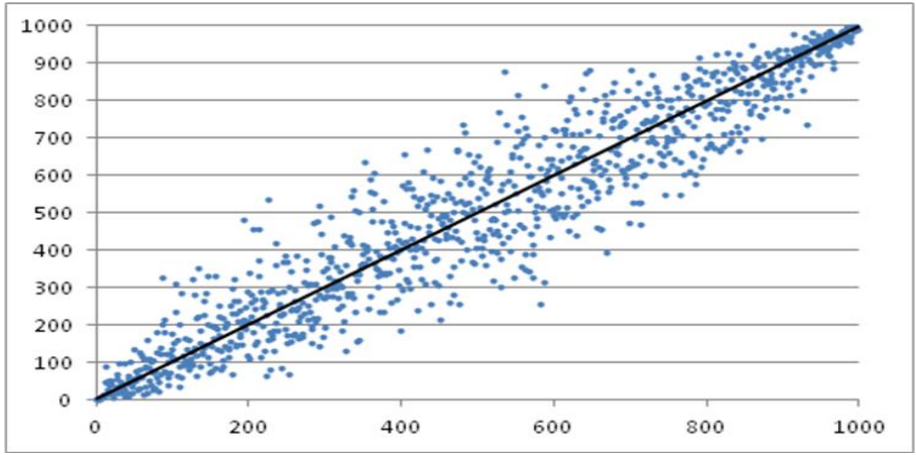

**Figure 8. Initial vs. perturbed ranks for α =0.7**

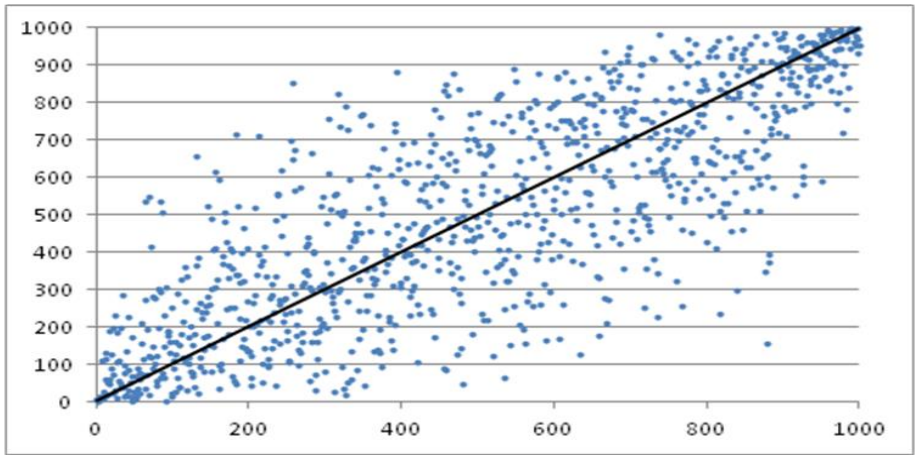

Swaps reinforce the fact that greater dissimilarity lower disclosure risk for the disseminated microdata perturbed by this method. Data points that are further away from the sample mean can be more easily identified due to two distinct problems: the classical issue of protection of the value recorded, plus a distance effect i.e. while perturbed an observation away from the mean could again face high disclosure risk by still remaining far from it. Swaps circumvent this additional problem. This mechanism is a (welcome) by-product of the present methodology.

Swaps in ranking, however, can also be a drawback, as data points swapping ranks will perturb the covariance with other variables. In fact, the lower $\alpha$ is, the lower the correlation between the original and the perturbed variable will be (Table 1); this will also imply higher perturbation of covariance with other variables.

**Table 1.   Correlation between original and perturbed variable coefficients for different similarity degrees**

| α | 0.999 | 0.95 | 0.9 | 0.8 | 0.7 |
|---|---|---|---|---|---|
| Correlation coefficient | 0.99 | 0.60 | 0.41 | 0.24 | 0.18 |

The Muralhidar-Sarathy hybrid generator outlined in the previous section automatically preserves some covariances, at least for the non-confidential variable used in the perturbation equation. However it remains mute for covariance with other confidential variables external to the equation. Through its similarity parameter, the multiplicative method presented here allows preserving the covariance with any other variables, but with a trade-off with the degree of securization that one wants to achieve in the disseminated data. This trade-off represents an inherent limit to the multiplicative masking structure. For example, one cannot adapt the perturbation process by introducing a non-confidential variable in order to preserve exactly some set of covariances: a necessary condition to do that would be that the non-confidential variable also follows a lognormal distribution. But a heavy-tailed non-confidential variable is a very unlikely configuration. In other cases, the use of perturbation method with any non-lognormal distribution would induce a distribution of the perturbed variable having a different functional neither exact nor closed form, or being too cumbersome an approximation to be tractable in a simple disclosure control environment (Laeven and al., 2005).

**Further remarks and conclusion**

When using statistical disclosure control techniques to generate perturbed data, analysis performed on the altered datasets should yield results that are identical or at least very close to those that would have been obtained using the original data. The assumption of normality in the distribution of the original variable and in the error term is a convenient way to achieve this objective. Unfortunately, many economic variables are distributed according to a heavy tailed, asymmetric form that makes the Gaussian framework limited. Moreover, as underlined in many recent studies (Piketty et al., 2003), fat tails are important for economic analysis as their impact could be substantial. Nonetheless, one has to note that data points generating heavy tailed distribution are often scarcely present in microdata sets, especially those coming from survey-based data (apart if specific oversampling procedures are used).

Two reasons account for this under-representation of high values. The first is simply due to the sampling scheme, as observations away from the mean are less likely to be observed in surveys. The second is that, as observations away from the mean face a higher disclosure risk than data points closer to it, control of these risk forces data producers to rely on top coding, i.e. values above a certain amount are automatically censored to that amount. As a result, a survey's skewedness is only a partial measure of the true population skewedness. In this case, one can still reasonably assume that normality is a sufficient assumption for surveys data perturbation, but further research will have to be conducted to determine the relative performances of these additive masking methods when the original data differ from a normal distribution.

The case of register-based microdata is quite different from that of surveys, as all the population is generally included. In this case, skewness is likely to occur very often, and our methodology will perform better than methods such as the Muralhidar-Sarathy hybrid generator. Moreover, as only heuristic rules are possible in practice for preserving covariances (one being, for example, choosing a degree of similarity between 0.99 and 0.95 that will protect observations away from the mean while preserving sufficiently the covariance), register-based data are favoured; due to their nature and the fact that they are not originally collected for analysis purpose, fewer variables are available than in a survey for covariance computations.

In conclusion, this paper has presented a simple technique that allows data producers to generate perturbed datasets according to a selectable degree of similarity when the underlying distribution is positively skewed, using the properties of lognormal distribution. The range of applications for this technique is potentially large, particularly when one is interested in the descriptive feature of a distribution. For example, this method avoids the use of interpolation for the computation of inequality as practiced in Atkinson et al., 2010. For a low value of the dissimilarity parameter, administrative records could easily made available as public use files. As argued by Sen and Foster (1997) in the case of income distribution: "The log-normal form gives good fits for many countries, though for high levels of income as such the best fits often seem to take the Pareto-form". This means that lognormal distributions, while useful and reasonable approximations, do not always conform to the heavy tails observed for some economic variables. However, transformations of distribution exhibiting more skewness than the lognormal form such as Pareto can only be achieved through approximation, rendering a disclosure control framework based on them intractable in practice. What is sure is that more research is needed in this field. Because of growing demand for microdata access, the simple methodology presented here could provide a useful starting point upon which more refined masking techniques preserving skewness could be built.